\documentclass[12pt]{article} 
\title{A Spacetime for SU(3)}  
\author{{\it Richard Shurtleff~}\thanks{affiliation and mailing 
address: Department of Applied Mathematics and Sciences, 
Wentworth Institute of Technology, 550 Huntington Avenue, 
Boston, MA, USA, ZIP 02115, telephone number: (617) 989-4338, fax 
number: (617) 989-4591 , e-mail address: shurtleffr@wit.edu}} 
\begin{document} 
          
\maketitle 

\begin{abstract} Rotations, boosts and translations in $8 + 1$ spacetime are developed based on the commutation and anticommutation relations of SU(3). The process follows a process that gives $3 + 1$ spacetime from SU(2).

\vspace{0.5cm}
 
\vspace{0.5cm}

\end{abstract}


\section{Introduction} \label{intro}

In a previous paper, hereafter `I',\cite{S1} the symmetries of spacetime were obtained from the algebra of the generators of SU(2). In this paper we apply the steps in I to SU(3) and determine the Lie algebra of a group of symmetries for `SU(3)-spacetime' with eight space dimensions and one time dimension. 

Here the generators $J$ of SU(3), elsewhere labeled `$F$', are called `angular momentum matrices.' Besides angular momentum matrices $J$, we define and discuss other matrices such as `boosts' $K$, `vector matrices' $V$ and `momentum matrices' $P$. Momentum matrices are vector matrices that commute. The  angular momentum matrices $J$ are considered as known in the many representations of SU(3), whereas the $K$s, $V$s, and $P$s are free to define as we wish.  

We select essential characteristics of the Poincar\'{e} algebra of spacetime to mimic in the new algebra for SU(3), the `SU(3)-Poincar\'{e} algebra'. In particular, the commutation relation $[P^{i},K^{j}]$ must be symmetric in $ij.$ Also $[P,J]$ and $[P,K]$ must be sums of $P$s.

Since the commutation relations of SU(3) are much the same as with SU(2), rotations in the eight dimensional space are obtained much as in I.

Boosts are more trouble than rotations. Boosts are directly tied to the anticommutation relations in the fundamental rep and the anticommutation relations of SU(3) include terms that do not occur with SU(2). The additional terms with SU(3) make a mess of the SU(3)-spacetime interval. The difference of the distance squared and the time squared, ${x^{i}}^{2}-{x^{t}}^{2},$ is not invariant under boosts in SU(3). 

There is, however, a cubic invariant that, when applied to coordinate differences, is invariant under rotations, boosts and translations.

Section 2 includes some properties of SU(3) and sets notation. Section 3 adapts the steps in I, leading from the fundamental 3-rep to the $3 \oplus 3$ = 6-rep to the $8 \oplus 1$ = 9-rep of SU(3). Along the way various $K$s and $P$s are defined so that by the end of Step 3 there is an SU(3)-Lorentz algebra of the $J_{(9)}$s and $K_{(9)}$s. Invariants are discussed in Step 4, including an expression for the boosted SU(3)-spacetime interval, a quadratic that is not invariant under boosts in SU(3).

To motivate investigating SU(3)-spacetime suppose one can say that the properties of four-dimensional spacetime accomodate the mechanical and electrodynamic behavior of `ordinary' matter composed of ordinary spin 1/2 particles, i.e. electrons, protons, and neutrons. And spin 1/2 is related to SU(2) because spin 1/2 occurs with two state systems whose unitary transformations are representations of SU(2).

However, we also observe particle behavior that involves the transformations of SU(3). It may be that we can guess the properties of SU(3)-spacetime by comparing SU(3) algebra with the algebras of SU(2) and four-dimensional spacetime symmetries. Then nine-dimensional SU(3)-spacetime geometry becomes available to help explain particle behavior.

\section{The Fundamental Rep} \label{FunRep}

Unitary $3\times3$ matrices with determinant equal to one form a group under matrix multiplication. A unitary  $3\times3$ matrix $D(\theta)$ can be written as a function of eight real parameters $\theta_{i},$ $i \in \{1,2,...,8\},$ with eight generators $J^{i}_{(3)},$
$ D_{(3)}(\theta)  \equiv$ $\exp{(i\theta_{i}J^{i}_{(3)})}.$ The $J$s are hermitian and traceless. The $J$s are called `angular momentum matrices' and the $\theta$ parameters are `angles'. Applying the `rotation matrix' $ D_{(3)}(\theta)$ to a 3-vector yields a rotated 3-vector. The fundamental rep is called the `triplet' rep or 3-rep of SU(3).\cite{GandG1,GandG2}

One set of $J^{i}_{(3)}$s is the set of Gell-Mann matrices,
\begin{equation} \label{Gell-Mann} J^{1}_{(3)} = \frac{1}{2} \pmatrix{0&&1&&0 \cr 1&&0&&0 \cr 0&&0&&0} \quad J^{2}_{(3)} = \frac{1}{2}\pmatrix{0&&-i&&0\cr i&&0&&0\cr0&&0&&0}  \quad J^{3}_{(3)} = \frac{1}{2}\pmatrix{1&&0&&0\cr0&&-1&&0\cr0&&0&&0} \quad,
\end{equation}
$$ J^{4}_{(3)} = \frac{1}{2}\pmatrix{0&&0&&1\cr0&&0&&0\cr1&&0&&0} \quad J^{5}_{(3)} = \frac{1}{2}\pmatrix{0&&0&&-i \cr 0&&0&&0\cr i&&0&&0}  \quad J^{6}_{(3)} = \frac{1}{2}\pmatrix{0&&0&&0\cr0&&0&&1\cr0&&1&&0} \quad, $$
$$ J^{7}_{(3)} = \frac{1}{2}\pmatrix{0&&0&&0\cr0&&0&&-i\cr0&&i&&0} \quad J^{8}_{(3)} = \frac{1}{2\sqrt{3}}\pmatrix{1&&0&&0 \cr 0&&1&&0\cr 0&&0&&-2}   \quad. $$
By inspection, these $J^{i}_{(3)}$s are hermitian and traceless. The label $J^{i}_{(3)}$ follows the notation in I; we do not use the usual notation $F^{i}$ for these matrices.

The commutators of the matrices satisfy
\begin{equation} \label{Comm1} [J^{i}_{(3)},J^{j}_{(3)}]  = i f^{ijk} J^{k}_{(3)}  \quad ,
 \end{equation}
where $i \in \{1,2,...,8\},$ the commutator is defined by $[J^{i},J^{j}] \equiv$ $J^{i}J^{j}-J^{j}J^{i},$ matrix multiplication is understood, the  $f^{ijk}$ are antisymmetric in $ijk.$ Summation over repeated indices is understood.

The anticommutators are
\begin{equation} \label{antiComm1} \{J^{i}_{(3)},J^{j}_{(3)}\}  = \frac{1}{3} \delta^{ij} {\mathbf{1}} + d^{ijk} J^{k}_{(3)}\quad,
 \end{equation}
where  ${\mathbf{1}}$ is the $3\times3$ unit matrix and the anticommutator is defined by $\{J^{i},J^{j}\} \equiv$ $J^{i}J^{j}+J^{j}J^{i},$ and $\delta^{ij}$ is the unit matrix, i.e. $\delta^{ij}$ = 0 for $i\neq j$ and $\delta^{ij}$ = 1 for $i = j.$ The $d^{ijk}$ are symmetric in $ijk$. 

The following identities hold for the antisymmetric coefficients $f$ and the symmetric coefficients $d,$
\begin{equation} \label{IDs}  f^{ijs}f^{skl}+f^{kjs}f^{sli}+f^{iks}f^{slj} = 0
\end{equation}
$$ f^{ijs}d^{skl}+f^{ljs}d^{ski}+f^{kjs}d^{sil} = 0 $$
$$ d^{ijs}d^{skl}+d^{ljs}d^{ski}+d^{lis}d^{skj} = \frac{1}{3} (\delta^{ki}\delta^{lj}+\delta^{kl}\delta^{ij}+\delta^{kj}\delta^{il}) \quad .$$

Another representation is the `antitriplet' or $\bar{3}$-rep. The basic matrices are the negative complex conjugate of the matrices $J^{i}_{(3)},$ $\bar{J}^{i}_{(\bar{3})}$ = $-(J^{i}_{(3)})^{\ast}$ = $-(J^{i}_{(3)})^{{\mathrm{T}}}.$  One can show that
\begin{equation} \label{Comm1BAR} [\bar{J}^{i}_{(\bar{3})},\bar{J}^{j}_{(\bar{3})}]  = i f^{ijk} \bar{J}^{k}_{(\bar{3})}  \quad
 \end{equation}
\begin{equation} \label{antiComm1BAR} \{\bar{J}^{i}_{(\bar{3})},\bar{J}^{j}_{(\bar{3})}\}  = \frac{1}{3} \delta^{ij} {\mathbf{1}} - d^{ijk} \bar{J}^{k}_{(\bar{3})}\quad.
 \end{equation}
{\it{Remark 0.1.}} Note that the commutation and anticommutation relations of the 3-rep and the $\bar{3}$-rep differ only in the sign of $d^{ijk}.$

\section{Four Steps to SU(3)-Spacetime} \label{4steps}
\noindent Step 1. {\it{Three dimensional vector matrices and boost matrices.}}

Start by defining nine vector matrices ${V^{\mu}_{(3)}}$, $\mu \in$ $\{1,2,...,9\},$  as follows,
\begin{equation} \label{Vs} V^{\mu}_{(3)} \equiv  \{c J^{i}_{(3)},c^{9}{\mathbf{1}} \}  \quad,
 \end{equation}
where $i \in$ $\{1,2,...,8\},$ $c$ and $c^{9}$ are arbitrary scalars and ${\mathbf{1}}$ is the $3\times 3$ unit matrix. Also define eight `boost matrices' ${K^{i}_{(3)}}$
\begin{equation} \label{Ks} K^{i}_{(3)}  \equiv + i J^{i}_{(3)}   \quad .
 \end{equation}
There is a $\pm$ sign ambiguity associated with the definition of $i$ = $\sqrt{-1},$ so we could just as well have chosen $K^{i}_{(3)}$ to be $-iJ^{i}_{(3)}.$ This is essential in setting up the $K$s in Step 2.

By the commutation relation (\ref{Comm1}) and the properties of the unit matrix, we see that the $V$s satisfy the commutation relation,
\begin{equation} \label{VJ} [V^{\mu}_{(3)},J^{j}_{(3)}]  =  i f^{\mu jk} V^{k}_{(3)}  \quad .
\end{equation}
We extend the range of indices to include nine by making $f^{9jk}$ vanish, $f^{9jk}$ = 0. 

By multiplying (\ref{Comm1}) by 1, $i$ and $i^{2},$ one sees that the generators $J_{(3)}$ and $K_{(3)}$ obey the commutation relations
 \begin{equation} \label{Comm2} [J^{i}_{(3)},J^{j}_{(3)}]  = i f^{ijk} J^{k}_{(3)}  \quad ; \quad [J^{i}_{(3)},K^{j}_{(3)}]  = i f^{ijk} K^{k}_{(3)}  \quad ; \quad [K^{i}_{(3)},K^{j}_{(3)}]  =  -i f^{ijk} J^{k}_{(3)} \quad.
 \end{equation}
These commutation relations form the basis for the `SU(3)-Lorentz Lie algebra' of angular momentum and boosts for SU(3). We include the commutation relations (\ref{VJ}) and (\ref{Comm2}) in the SU(3)-Poincar\'{e} algebra.

The SU(3)-Poincar\'{e} algebra mimics some of the characteristics of the Poincar\'{e} algebra. We require that $[V^{i},K^{j}]$ = $ik^{ij\mu}V^{\mu}$ with $ij$-symmetric coefficients $k^{ij\mu}$ = $k^{ji\mu}.$ But this cannot be true for $V^{i}_{(3)}$ and $K^{i}_{(3)}$ because both of them are  multiples of $J^{i}_{(3)}.$ The 3-rep fails to give a complete basis for the SU(3)-Poincar\'{e} algebra.

\pagebreak 
\noindent Step 2. {\it{The $(3\oplus {3})$ rep of SU(3) gives the complete set of basic SU(3)-Poincar\'{e} commutation relations.}}
\vspace{0.01cm}

The trick to getting $[V^{i},K^{j}]$ symmetric in $ij$ is to make the commutators $[V^{i},K^{j}]$ depend on anticommutators. To do this we work with the $(3\oplus 3)$ = 6-rep of SU(3).

Define carefully $6\times6$ matrices $K_{(6)}$, and $V_{(6)}$ to accompany the generators $J_{(6)}$ of the 6-rep. We have
 \begin{equation} \label{anti1} J^{i}_{(6)} = \pmatrix{J^{i}_{(3)} && 0 \cr 0 && J^{i}_{(3)}}  \quad ; \quad K^{i}_{(6)} = \pmatrix{+K^{i}_{(3)} && 0 \cr 0 && -K^{i}_{(3)}}   \quad ; \quad  V^{\mu}_{(6)} = \pmatrix{0 && V^{\mu}_{+(3)} \cr V^{\mu}_{-(3)} && 0} \quad,
 \end{equation}
where
 \begin{equation} \label{anti2} V^{\mu}_{+(3)} \equiv  \{c_{+} J^{i}_{(3)},c^{9}_{+}{\mathbf{1}} \}  \quad {\mathrm{;}}\quad  V^{\mu}_{-(3)} \equiv  \{c_{-} J^{i}_{(3)},c^{9}_{-}{\mathbf{1}} \}  \quad .
 \end{equation}
The constants $c_{+},$ $c^{9}_{+},$ $c_{-}$ and $c^{9}_{-}$ are possibly different choices for the $c$s in (\ref{Vs}).

One can verify the SU(3)-Poincar\'{e} commutation relations from Step 1, 
 \begin{equation} \label{Comm3a} [V^{\mu}_{(6)},J^{j}_{(6)}]  =  i f^{\mu jk} V^{k}_{(6)} \quad
 \end{equation}
and 
 \begin{equation} \label{Comm2a} [J^{i}_{(6)},J^{j}_{(6)}]  = i f^{ijk} J^{k}_{(6)}  \quad ; \quad [J^{i}_{(6)},K^{j}_{(6)}]  = i f^{ijk} K^{k}_{(6)}  \quad ; \quad [K^{i}_{(6)},K^{j}_{(6)}]  =  -i f^{ijk} J^{k}_{(6)} \quad.
\end{equation}
Thus the $J$s, $V$s and $K$s of the 6-rep of SU(3) satisfy the $[J,J],$ $[J,K],$ $[K,K],$ $[V,J]$ commutation relations of the 3-rep in Step 1.

Now we calculate the commutators $[V^{i}_{(6)},K^{j}_{(6)}]$ and require that they be sums of $V_{(6)}$s and that they be symmetric in $ij.$ One finds by (\ref{antiComm1}), (\ref{Ks}), (\ref{anti1}) and (\ref{anti2}) that
 \begin{equation} \label{Comm4} [V^{i}_{(6)},K^{j}_{(6)}] =  \pmatrix{0&&-ic_{+}\{J^{i}_{(3)},J^{j}_{(3)}\} \cr +ic_{-}\{J^{i}_{(3)},J^{j}_{(3)}\}  &&0} = \end{equation}  
$$ = -i \sqrt{\frac{2}{3}}\alpha\delta_{ij} V^{9}_{(6)} -i d^{ijk}\beta V^{k}_{(6)} + i\Delta^{ij} \quad ,$$
where we include a factor of $\sqrt{2/3}$ with the coefficient $\alpha$ for convenience below and $\Delta$ is given by 
$$\Delta^{ij} = \frac{ \delta_{ij}}{3}\pmatrix{0&& (\sqrt{6}\alpha c^{9}_{+} -c_{+})\, {\mathbf{1}}\cr (\sqrt{6}\alpha c^{9}_{-} +c_{-})\, {\mathbf{1}} &&0} +   d^{ijk}\pmatrix{0&& c_{+}(\beta-1) J^{k}_{(3)}\cr c_{-}(\beta+1) J^{k}_{(3)}  &&0}$$ 
Thus, by making $\Delta^{ij}$ vanish, we can have the $[V^{i}_{(6)},K^{j}_{(6)}]$ be sums of $V_{(6)}$s and also be symmetric in $ij.$ 

 From $\Delta^{ij}$ = 0 and the linear independence of the set $\{J^{k}_{(3)},{\mathbf{1}}\}$ it follows that 
 \begin{equation} \label{alphabeta} \sqrt{6}\alpha c^{9}_{+} = c_{+}\quad ; \quad \sqrt{6}\alpha c^{9}_{-} = -c_{-}\quad ; \quad c_{+}(\beta - 1) = 0\quad ; \quad c_{-}(\beta + 1) = 0\quad .
 \end{equation}
Now, for nonzero $\alpha,$ $\alpha \neq$ 0, we have either
 \begin{equation} \label{beta} \beta = +1  \quad ; \quad  c_{-} = 0   \quad ; \quad    c^{9}_{-} = 0    \quad ; \quad c^{9}_{+} = +\frac{c_{+}}{\sqrt{6}\alpha}  \quad \end{equation} or we have $$\beta = -1  \quad ; \quad  c_{+} = 0  \quad ; \quad    c^{9}_{+} = 0    \quad ; \quad c^{9}_{-} = -\frac{c_{-}}{\sqrt{6}\alpha}  \quad  $$
 or, and this is a case that we ignore, $\beta$ is some other number and $c_{+}$ = $c_{-}$ = 0, the case with vanishing $V$s.

By (\ref{anti1}), when either $c_{+}$ or $c_{-}$ vanish, the vector matrices commute, $[V^{\mu}_{(6)},V^{\nu}_{(6)}]$ = 0. Commuting vector matrices are momentum matrices, so we have two sets of momentum matrices. 

One momentum `$P^{(+)}$' goes with $\beta$ = 1 and $c_{+} \neq$ 0 and the other momentum `$P^{(-)}$' applies when $\beta$ = $-1$ and $c_{-} \neq$ 0. With $V \rightarrow$ $P$ in (\ref{anti1}), we have
\begin{equation} \label{P+} P^{(+)\,i}_{(6)} \equiv \pmatrix{0 && c_{+}J^{i}_{(3)} \cr 0 && 0}  \quad {\mathrm{;}}\quad  P^{(+)\,9}_{(6)} \equiv \pmatrix{0 && c_{+}{\mathbf{1}}/(\sqrt{6}\alpha) \cr 0 && 0} \quad ;
 \end{equation}
\begin{equation} \label{P-} P^{(-)\,i}_{(6)} \equiv \pmatrix{0 && 0\cr c_{-}J^{i}_{(3)}  && 0}  \quad {\mathrm{;}}\quad  P^{(-)\,9}_{(6)} \equiv \pmatrix{0 && 0 \cr -c_{-}{\mathbf{1}}/(\sqrt{6}\alpha) && 0} \quad .
 \end{equation}
By  (\ref{anti2}) - (\ref{P-}), the matrices $J^{i}_{(6)},$ $P^{(\pm) \, j}_{(6)}$ and $K^{k}_{(6)}$ obey the following commutation relations
$$ [J^{i},J^{j}]  = i f^{ijk} J^{k}  \quad ; \quad [J^{i},K^{j}]  = i f^{ijk} K^{k}  \quad ; \quad [K^{i},K^{j}]  =  -i f^{ijk} J^{k} \quad
$$
\begin{equation} \label{Pcomm1} [P^{(\pm)\,i},K^{j}] = -i (\sqrt{\frac{2}{3}}\delta_{ij}\alpha P^{(\pm)\,9} \pm  d^{ijk} P^{(\pm)\,k}) \quad ; \quad 
[\alpha P^{(\pm)\,9},K^{j}]  =  - i \sqrt{\frac{2}{3}}  P^{(\pm)\,j}  \quad ;\end{equation}
$$ [P^{(\pm)\,\mu},J^{j}]  =  i f^{\mu jk} P^{(\pm)\,k}  \quad ; \quad  [P^{(\pm)\,\mu},P^{(\pm)\,\nu}] = 0 \quad .$$ 
We take these commutation relations to be the basis of the `SU(3)-Poincar\'{e} Lie algebra'. 

Since $\alpha$ occurs with every time component, index = 9, it determines the ratio of the eight spatial components to the time. For ordinary spacetime, $\alpha$ is the speed of light. Just as in spacetime, it is convenient to set the quantity $\alpha$ to unity,
\begin{equation} \label{alpha} \alpha = 1  \quad .
\end{equation}
At this point, we have no concept in SU(3)-spacetime like the `speed of light' in spacetime. The most one can say is that the unit for distance in the eight dimensional space and the unit for time are equal. 

{\it{Remark 2.1.}} The two SU(3) spacetimes with $\alpha$ = $\pm 1$ are related by a time inversion or a parity inversion. 

{\it{Remark 2.2.}} The SU(3)-Poincar\'{e} algebras for $P^{(+)}$ and $P^{(-)}$ differ by the sign of the symmetric coefficients $d^{ijk}.$ By Remark 0.1, replacing $P^{(\pm)}$ by $P^{(\mp)}$ is equivalent to replacing the 3-rep with the $\bar{3}$-rep in Steps 1 and 2. 

{\it{Remark 2.3.}} Since the $P^{(\pm)\,\mu}$s commute and the commutators $[P^{\pm},X]$ are sums of the $P^{\pm}$s for any $X$ = $J,$ $K,$ or $P^{\pm},$ the $P^{\pm}$s form an {\it{abelian ideal}}.

\vspace{0.4cm}
\noindent Step 3. {\it{In this step each set, one set for `$+$' one for `$-$', of nine momentum matrices $P^{(\pm)\,\mu}_{(6)},$ $\mu \in$ $\{1,2,...,9\},$ acts as a catalyst, carrying the SU(3)-Lorentz algebra of the $J^{i}_{(6)}$s and $K^{j}_{(6)}$s from the $3\oplus3$ = $6$-rep to the $8\oplus1$ = $9$-rep of SU(3). The momentum matrices themselves are not carried to nine-dimensional SU(3)-spacetime by the procedure.}}
\vspace{0.4cm}

Three facts underly the process in this step. First, the Lorentz Lie algebra of $J$s and $K$s is closed. Second, the commutators $[P,J]$ and $[P,K]$ are sums of $P$s. Third, the nonzero blocks of the matrices for the $P^{(\pm)\,\mu}_{(6)}$s are off-diagonal.

Choose a matrix $A^{i}$ and a matrix $B^{j}$ from the collection of $J^{i}_{(6)}$s and $K^{i}_{(6)}$s. One can write the commutation relation as follows,
\begin{equation} \label{ABC} [A^{i},B^{j}] = s^{ijk}_{(AB)} C^{k}   \quad,
\end{equation}
where the $C^{k}$s are  $J^{k}_{(6)}$s or $K^{k}_{(6)}$s and the coefficients $s^{ijk}_{(AB)}$ are found in (\ref{Pcomm1}) for the particular choice of $A^{i}$ and $B^{j}$. 

By (\ref{Pcomm1}), we have
\begin{equation} \label{VABC} [P^{(\pm)\,\mu}_{(6)},A^{i}] = a^{\mu i \nu} P^{(\pm)\,\nu}_{(6)} \quad ; \quad  [P^{(\pm)\,\mu}_{(6)},B^{j}] = b^{\mu j \nu} P^{(\pm)\,\nu}_{(6)} \quad ; \quad  [P^{(\pm)\,\mu}_{(6)},C^{k}] = c^{\mu k \nu} P^{(\pm)\,\nu}_{(6)}  \quad,
\end{equation}
where the coefficients $a$, $b$ and $c$ can be read from the equations for the particular choice of $A,B,C$ in (\ref{ABC}). From (\ref{VABC}), one can show that 
$$[P^{(\pm)\,\rho}_{(6)},[A^{i},B^{k}]] =     \left( a^{\rho i \mu} b^{\mu k \nu} -  b^{\rho k \mu} a^{\mu i \nu} \right) P^{(\pm)\,\nu}_{(6)}  \quad.
$$
Thus the momentum matrices transfer the commutator of $A$ and $B$ to a commutator of $a$ and $b.$

Let $a^{i}$ be the $9\times9$ matrix with a $\rho \mu$th component equal to $a^{\rho i \mu},$ $\mu,\rho \in$ $\{1,2,...,9\}.$ Similarly define $b^{k}$ and $c^{n}.$ With (\ref{ABC}), the last equation implies that 
\begin{equation} \label{abP} \left[ a^{i},b^{k}\right]_{\rho \nu} P^{(\pm)\,\nu}_{(6)} = s^{ikn}_{(AB)}c^{n}_{\rho \nu} P^{(\pm)\,\nu}_{(6)} \quad.
\end{equation}
One cannot simply cancel the $P^{(\pm)\,\nu}_{(6)}$s  because there is a sum over $\nu.$ And the nine $P^{(\pm)\,\nu}_{(6)}$s, $\nu \in$ $\{1,2,...,9\}$ do not form a linearly independent set of 36-component $6\times6$ matrices. 

But the $P^{(\pm)\,\nu}_{(6)}$ are off-diagonal in this representation; see (\ref{P+}) and (\ref{P-}). And the nonzero 3-dimensional blocks ${P^{(+)\,\nu}_{+}}$ and ${P^{(-)\,\nu}_{-}},$ each proportional to $\{ J^{i}_{(3)}, {\mathbf{1}}/3 \},$ do form  linearly independent sets of nine-component $3\times3$ matrices. Then the preceding equation (\ref{abP}) reduces to two equations, one in the 12-block (subscript +) and one in the 21-block (subscript $-$). The 12- and 21-blocks are 3-dimensional matrices and, by the linear independence of the nine $3\times3$ matrices ${P^{(\pm)\,\nu}_{+}}$ and ${P^{(\pm)\,\nu}_{-}},$ it follows that that 
\begin{equation} \label{theorem3} \left[ a^{i},b^{k}\right] = s^{ikn}_{(AB)}c^{n} \quad .
\end{equation}
The coefficients $a^{i},$ $b^{k}$ and $c^{n}$ obey the same commutation relation (\ref{ABC}) as $A^{i},$ $B^{k}$ and $C^{n}.$ 

By (\ref{theorem3}), the coefficients in (\ref{Pcomm1}) form angular momentum matrices ${J^{i}_{(9)}}_{\rho \mu}$ and boost matrices ${K^{i}_{(9)}}_{\rho \mu},$
\begin{equation} \label{JK9} {J^{i}_{(9)}}_{\mu \nu}  =  if^{\mu i \nu} \quad ; \quad {K^{\pm}_{(9)}}^{i}_{\mu \nu}  =  -i \left[\sqrt{\frac{2}{3}}\left(\delta^{i}_{\mu} \delta^{9}_{\nu} +\delta^{i}_{\nu} \delta^{9}_{\mu}\right) \pm  d^{\mu i \nu}\right] \quad,
\end{equation}
where time components of $f^{\mu i \nu}$ and $d^{\mu i \nu}$ vanish, i.e. $f^{9\alpha \beta}$ = 0, $d^{9\alpha \beta}$ = 0 with permutations. 

By (\ref{theorem3}) or directly from (\ref{JK9}), these ${J^{i}_{(9)}}$s and ${K^{\pm}}^{i}_{(9)}$s satisfy the SU(3)-Lorentz Lie algebra,
 \begin{equation} \label{Comm2b} [J^{i}_{(9)},J^{j}_{(9)}]  = i f^{ijk} J^{k}_{(9)}  \quad ; \quad [J^{i}_{(9)},{K^{\pm}}^{j}_{(9)}]  = i f^{ijk} {K^{\pm}}^{k}_{(9)} \quad ; \quad [{K^{\pm}}^{i}_{(9)},{K^{\pm}}^{j}_{(9)}]  =  -i f^{ijk} J^{k}_{(9)} \quad, 
\end{equation} 
where `$(9)$' is short for $(8\oplus 1)$ and indicates the `9-vector' rep. The ${J^{i}_{(9)}}$s generate rotations in the eight-dimensional space of SU(3)-spacetime. The ${K^{i}_{(9)}}$s generate boosts in nine dimensional SU(3)-spacetime.

{\it{Remark 3.1.}} By tracing the origins of the angular momentum matrices ${J^{i}_{(9)}}$ in (\ref{JK9}) through (\ref{VABC})  to (\ref{Comm3a}) to (\ref{Comm1}), one sees that the ${J^{i}_{(9)}}$s are based on the structure constants $f^{ijk}$ of the commutation relations, $[J^{i}_{(3)},J^{j}_{(3)}]$ = $i f^{ijk} J^{k}_{(3)},$ in the fundamental 3-rep of SU(3) in (\ref{Comm1}). 

{\it{Remark 3.2.}} Tracing the origins of the boost matrices ${K^{i}_{(9)}}$ (\ref{JK9}), we follow the ${K^{i}_{(9)}}$s from (\ref{VABC}) to  (\ref{Pcomm1}) with (\ref{Comm3a}) back to (\ref{antiComm1}). One sees that the $\delta$s and $d^{ijk}$ terms in the ${K^{i}_{(9)}}$s evolve from the delta function and $d^{ijk}$ term in the anticommutator relation (\ref{antiComm1}) in the fundamental 3-rep, $\{J^{i}_{(3)},J^{j}_{(3)}\}$ = $\delta^{ij}/3 + d^{ijk}J^{j}_{(3)}.$   

\pagebreak 
\noindent Step 4. {\it{Invariants. The square of the distance in eight-dimensional space ${x^{i}}^{2},$  the quantity $d^{ijk}x^{i}y^{j}z^{k},$ and time are invariant under rotations. There is no quadratic invariant like the spacetime interval. Instead invariance under rotations and boosts is found for a}}  cubic {\it{quantity. }}
\vspace{0.3cm}

We use the SU(3)-Lorentz transformation $D^{\pm}_{(9)}(\theta,\phi)$ = $\exp{(i\phi_{i}{K^{\pm}}^{i}_{(9)})}\exp{(i\theta_{i}J^{i}_{(9)})},$ the 9-vector transformation matrix for a rotation through angle $\theta$ = $\{\theta_{1},\theta_{2},...,\theta_{8}\}$ followed by a boost through $\phi$ = $\{\phi_{1},\phi_{2},...\phi_{8}\}.$

For an SU(3)-Lorentz transformation followed by a translation applied to the coordinates $x^{\mu}_{0}$ of some point in nine dimensional SU(3)-spacetime, we use the following device,
\begin{equation} \label{RBT1} 
  \pmatrix{[D_{(9)}(\theta,\phi)]^{\mu}_{\nu}&&a^{\mu} \cr 0 && 1} \pmatrix{x^{\nu}_{0} \cr 1} = \pmatrix{x^{\prime \, \mu}_{0} +a^{\mu} \cr 1}\quad .
 \end{equation}
 The prime indicates $x_{0}$ has been rotated then boosted, $x^{\prime}_{0}$ = $D_{(4)}(\theta,\phi)\,x_{0}. $

One can show the invariance under rotations of the square of the distance, the cubic quantity $d^{ijk}x^{i}x^{j}x^{k},$ and the time $x^{9}.$ One has 
 \begin{equation} \label{x2ROTa}  {{x^{ \prime}}^{ i}}^{2} = {x^{i}}^{2} \quad ; \quad d^{ijk}{{x^{ \prime}}^{ i}}{{x^{ \prime}}^{ j}}{{x^{ \prime}}^{k}} = d^{ijk}x^{i}x^{j}x^{k} \quad ; \quad {{x^{ \prime}}^{ 9}} = {x^{9}} \quad,
 \end{equation}
where the prime indicates the rotated 9-vector ${x^{ \prime}}$ = $D^{+}_{(9)}(\theta,0)x$  = $D^{-}_{(9)}(\theta,0)x.$ The square of the distance is invariant because the rotation generators ${J^{i}_{(9)}}_{\mu \nu}$  are antisymmetric in $\mu\nu.$ The cubic quantity invariance is a consequence of an $f,d$ identity (\ref{IDs}) and time is invariant because the time components of $J^{i}$ vanish. 

Turn now to boosts, ${x^{ \prime \prime}}_{(\pm)}$ = $D^{\pm}_{(9)}(0,\phi)x,$ where $D^{\pm}_{(9)}$ is generated with $K^{\pm \, i}_{(9)}.$ There is no quadratic invariant for boosts. The problem is the symmetric coefficients $d^{ijk}.$ One finds that
 \begin{equation} \label{x2Boost} {{x^{\prime \prime \, i}_{(\pm)}}}^{2} - {{x^{\prime \prime \, 9}_{(\pm)}}}^{2} = {x^{i}}^{2} - {x^{9}}^{2} \pm 2 \phi_{m}d^{jmk}x^{j}x^{k} \quad ,
 \end{equation}
where $\phi$ is small, $\mid \phi_{m} \mid \ll$ 1, and terms of order $\phi^{2}$ are dropped. 

{\it{Remark 4.1.}} The $d$-dependence of boosts enters with the boost generators ${K^{\pm}}^{j}_{(9)}$ in (\ref{JK9}) as a consequence of the fundamental anticommutation relation, (\ref{antiComm1}). See Remark 3.2. There are no symmetric coefficients for SU(2), $d^{ijk} \rightarrow 0$, and with SU(2)-spacetime (four dimensional spacetime) the quadratic ${x^{i}}^{2} - {x^{9}}^{2}$ is invariant under boosts. See I. With nonzero $d^{ijk},$ $d^{ijk} \neq 0,$ the SU(3)-spacetime interval is not an invariant.

Define the two cubic quantities, one for $+$ one for $-$, $I^{\pm}(x)$ = $g^{\pm}xxx$ by
  \begin{equation} \label{gxxx}  I^{\pm}(x) = g^{\pm}_{\alpha \beta \gamma} x^{\alpha} x^{\beta} x^{\gamma} = \mp \sqrt{\frac{3}{2}} d_{ijk} x^{i}x^{j}x^{k} + \frac{3}{2} {x^{i}}^{2} x^{9} - {x^{9}}^3 \quad,
 \end{equation}
where the signs are keyed to the $\pm$ in the boost transformation $D^{\pm}_{(9)}(0,\phi).$ 

The cubics $I^{\pm}(x)$ are made with rotation invariants ${x^{i}}^{2}$, $d^{ijk}x^{i}x^{j}x^{k},$ and $x^{9}.$ Therefore the $I^{\pm}(x)$s are rotation invariants. One can show that $I^{\pm}(x)$s are also invariant under boosts. When $x$ is the coordinate difference of two 9-vectors, $x$ itself is unchanged by a translation. It follows that, assuming $x$ is proportional to a coordinate difference,  the quantities $I^{\pm}(x)$ are invariant under rotations, boosts and translations.

\appendix


\section{Exercises and Problems} \label{Pb}

\noindent Exercises and problems are provided online.

\vspace{0.3cm}
\noindent 1. Scalar products in eight dimensional space. Show that $x^{i}y^{i}$ and $d_{ijk} x^{i}y^{j}z^{k}$  are invariant under rotations. (Hint: try (\ref{x2ROTa}) with $x \rightarrow$ $x+y$ and $x \rightarrow$ $x+y+z.$)

\vspace{0.3cm}
\noindent 2. A scalar product in nine dimensional SU(3)-spacetime. Define $I^{\pm}(x,y,z)$ by \begin{equation} \label{gxyz}  I^{\pm}(x,y,z) = g^{\pm}_{\alpha \beta \gamma} x^{\alpha} y^{\beta} z^{\gamma} = \left[ \frac{1}{2}\left( \delta_{ij} \delta^{9}_{k} + \delta_{ki} \delta^{9}_{j}+ \delta_{jk} \delta^{9}_{i}\right) \mp \sqrt{\frac{3}{2}} d_{ijk}\right]x^{i}y^{j}z^{k} -x^{9}y^{9}z^{9} \quad,
 \end{equation}
Show that $I^{\pm}(x,y,z)$ is invariant under rotations and boosts.

\vspace{0.3cm}
\noindent 3. Find the generators $J^{i}_{(10)},$ ${K^{\pm \, i}_{(10)}},$ $P^{\mu}_{(10)}$ of the SU(3)-Poincar\'{e} transformation in (\ref{RBT1}). Show the matrices obey the SU(3)-Poincar\'{e} algebra (\ref{Pcomm1}). What is interesting about the signs?

\vspace{0.3cm}
\noindent 4. Given the SU(3)-Lorentz transformation for a rotation followed by a boost $\Lambda(\theta,\phi)$ = $\exp{(i\phi_{i}{K^{\pm}_{(9)}}^{i})}\exp{(i\theta_{i}J^{i}_{(9)})}$ and the general SU(3)-Poincar\'{e} rep transformation $D(\theta,\phi)$ = $\exp{(i\phi_{i}K^{i})}\exp{(i\theta_{i}J^{i})}$ with  $J$s, $K$s and $V$s obeying (\ref{Pcomm1}), show that $DV^{\mu}D^{-1} = \Lambda^{\mu}_{\nu}V^{\nu}.$


\begin{thebibliography}{9}

\bibitem{S1} R. Shurtleff,  {\it{Spacetime is for SU(2)}}, 	http://arxiv.org/abs/1001.1425 , article in the on-line physics arxiv.



\bibitem{GandG1} See, for example, W. Greiner and B. M\"{u}ller, {\it{Quantum Mechanics - Symmetries}}, 2nd ed. (Springer-Verlag, Berlin, 1994), Chapter 7 and references therein.

\bibitem{GandG2} See, for example, S. Gasiorowicz, {\it{Elementary Particle Physics}} (Wiley, New York, 1967).


\end{thebibliography}
\end{document}